\documentclass[reprint,superscriptaddress,twocolumn,%aip,
apl,
showpacs,floatfix,shortbibliography]{revtex4-1}
\usepackage{graphicx}
\usepackage{amsbsy,amssymb,amsmath,bm,ulem}
\usepackage{hyperref}% add hypertext capabilities
\usepackage[mathlines]{lineno}% Enable numbering of text and display math
\usepackage{csquotes} % for left quotation marks
\usepackage{xcolor}  % for colored text.

\begin{document}

\preprint{APS/123-QED}

\title{Measurement of critical current flow and connectivity in systems of  joined square superconducting plates}

\author{F. Colauto}
\affiliation{Departamento de F\'{i}sica, Universidade Federal 
de S\~{a}o Carlos, 13565-905, S\~{a}o Carlos, SP, Brazil}

\author{D. Carmo}
\affiliation{Departamento de F\'{i}sica, Universidade Federal 
de S\~{a}o Carlos, 13565-905, S\~{a}o Carlos, SP, Brazil}
\affiliation{Laborat\'{o}rio Nacional de Luz S\'{i}ncrotron, Centro Nacional de Pesquisa em Energia e Materiais, 13083-100, Campinas, SP, Brazil}

\author{A. M. H. de Andrade}
\affiliation{Instituto de F\'{i}sica, Universidade Federal do 
Rio Grande do Sul, 91501-970, Porto Alegre, RS, Brazil}

\author{A.~A.~ M.~Oliveira}
\affiliation{Instituto Federal de Educa\c{c}\~{a}o, Ci\^{e}ncia e Tecnologia 
de S\~{a}o Paulo, 13565-905, S\~{a}o Carlos, SP, Brazil}

\author{W. A. Ortiz}
\affiliation{Departamento de F\'{i}sica, Universidade Federal 
de S\~{a}o Carlos, 13565-905, S\~{a}o Carlos, SP, Brazil}

\author{Y. M. Galperin}
\affiliation{Department of Physics, University of Oslo, P. O. Box 1048 
Blindern, 0316 Oslo, Norway}

\author{T. H. Johansen}
\affiliation{Department of Physics, University of Oslo, P. O. Box 1048 
Blindern, 0316 Oslo, Norway}
\affiliation{Institute for Superconducting and Electronic Materials, 
University of Wollongong, Northfields Avenue, Wollongong, 
NSW 2522, Australia}

\date{\today}

\begin{abstract}
A method to measure the electrical connectivity between square superconducting plates joined by weak link interfaces is presented.
It is based on observation of lines where the flow of critical current abruptly changes 
direction due to the presence of weak links, and the confinement created by the shape of the sample.
The method is demonstrated using  magneto-optical imaging (MOI) of systems consisting of up 
to $2 \times 2$ plates joined to form a larger square.
Common features are found in the current flow patterns, which allow to measure the electrical connectivity between the plates by observing an angle between pairs of lines indicating where the current abruptly changes flow direction, so-called discontinuity, or d-lines.
The samples used in this study are Nb films with weak links created by focused ion beam machining.  
\end{abstract}

\maketitle

\section{Introduction}
State-of-the-art bulk superconductors are today able to trap magnetic flux at such high densities, and density gradients, that the corresponding electrical currents create magnetic fields well above
 10~T~\cite{durrell_bulk_2018}, i.e., far larger than fields from conventional permanent magnets.
Another distinct property of fully magnetized  superconductors is that their trapped field increases not only with the thickness in the magnetization direction, but also  with their lateral size.
Both these properties are illustrated by the case of a circular superconducting disk of thickness $t$, and radius $R$.
When saturated throughout its volume by a circulating  critical current of density $J_c$, the disk creates at the center of its flat surfaces a perpendicular field of magnitude given by~\cite{johansen_flux-pinning-induced_2000},

\begin{equation*}
B = \frac{\mu_0 J_c t}{2} \, \ln[ R/t+ \sqrt{(R/t)^2+1  } ] \, ,
\end{equation*}
where $ \mu_0$ is the permeability of vacuum.

While  permanent magnets and superconductors shaped as circular disks are commonly available, they obviously fail to allow  assembly into  compact  plane-filling arrangements.
For that,  elements shaped as the unit cell of a 2D Bravais lattice will work.
A most versatile alternative is then to use  square elements.

When combining superconducting elements the crucial issue is to maximize the flow of electrical current through the joining interfaces.
To monitor that, methods to detect flow patterns of supercurrents are needed.
Several approaches have been applied to measure the electrical connectivity between joined superconductors.
Kordyuk~\textit{et~al.} suggested a method based on recording the levitation force on a small magnet while being displaced over a surface area that includes the joining interface~\cite{kordyuk_simple_2001}. 
Another approach is to scan such areas with a Hall probe, or an array of them~\cite{bending_local_1999}.

In the present work magneto-optical imaging (MOI) was applied to observe patterns of flux penetration in square superconducting plates connected by weak links.
Focus is set on the highly visible lines in the flux density distribution indicating where the critical current makes abrupt changes in the flow direction.  
It is demonstrated how the configuration of these so-called discontinuity, or d-lines, allows to reconstruct the full pattern of critical current flow in systems of sizes up to $2 \times 2$ square plates joined to form a larger square. 
It is also shown that the d-line pattern allows to make quantitative measurements of the electrical connectivity of the joining interfaces.

\section{Experimental}
Superconducting films of Nb were deposited on Si (100) substrates by magnetron sputtering in a UHV chamber with base pressure below $2 \times 10^{-8}$~Torr. 
The films studied in this work are 300 nm thick, and optical lithography was used to shape them as square and rectangular samples. 
The critical temperature of the samples is $T_c$~=~9.1~K with a sharp superconducting transition, $\Delta T_c$ = 0.1~K, obtained by AC susceptibility. Separate MOI measurements of unpattered samples showed that the Nb films have a critical current density of the order of $10^6$~A/cm$^2$.

Two square films were prepared with sides of 2.5~mm. 
In one of them, a pair of perpendicular grooves were added using focussed ion beam machining.
This created a sample consisting of 4 equal squares connected by weak links.
The other square film was kept plain to serve as a reference sample. 
In a rectangular film of size $2.5 \times 5.0$~mm$^2$ one groove was added to create a sample consisting of 2 equal squares connected by a weak link.
The equipment used to make the grooves was a JEOL JIB-4500 Multi-Beam System SEM-FIB with probe current of 5~nA.
This provides a gallium ion dose of 0.1 nC/$\mu$m$^2$.

The MOI observations were carried out using a setup where a Bi-substituted ferrite garnet film (FGF) with in-plane magnetization was placed directly on the sample, where it serves as a Faraday-active flux density sensor.
The FGF was made by liquid epitaxial growth on gadolinium gallium garnet substrates, as described elsewhere~\cite{helseth_faraday_2002}. When viewed in a polarized light  microscope with crossed polarizers, MOI allows to follow in real-time the global penetration of magnetic flux into the superconducting samples~\cite{vlasko-vlasov_magneto-optical_1999, altshuler_2004, yurchenko_dendritic_2009}.

The optical cryostat used  in this work is an Oxford MicrostatHe-R, where the sample is mounted on a cold finger at the end of a bayonet. 
By inserting the bayonet into the cryostat, the sample is placed below an optical window, thus allowing magneto-optical imaging (MOI) to be performed.
Magneto-optical images of the samples were obtained using a BX-LRA-2 Olympus microscope equipped with a U-PO3 polarizer, a U-AN360-3 analyzer, and a 5x MPlanFL N objective.
The images were recorded by a digital CCD camera Retiga 4000R Q-imaging. 
Magnetic fields were applied perpendicular to the samples using a pair of coils in a Helmholtz configuration.

In the magneto-optical images presented in this paper, the brightness represents the local density of perpendicular magnetic flux.
The experiments were all carried out at $T =$ 7.0~K, a temperature sufficiently high to avoid intermittent flux dynamics due to thermomagnetic instabilities~\cite{yurchenko_dendritic_2009} in Nb samples~\cite{colauto_boundaries_2008}.

\section{Results}

\subsection{One square}
\begin{figure}[t]
  \centering
  \includegraphics[width=8.5 cm]{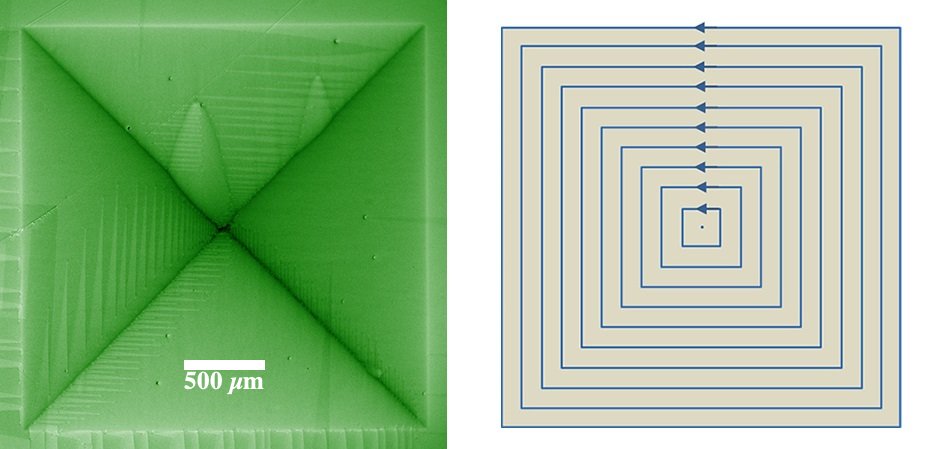} 
  \caption{
  Left: Magneto-optical image of the flux distribution in a square Nb film exposed to a perpendicular magnetic field of 220~Oe, creating full flux penetration.
  The dark lines are regions of strong shielding.
  The contour of the sample is seen as a bright square. 
  Zig-zag patterns are related to domain walls in the FGF, irrelevant for the present work. 
  Right: Schematic of the critical current streamlines in a square superconducting plate in a fully penetrated state created by a perpendicular magnetic field applied.}
  \label{fig:Fig1}
\end{figure}

As a reference, let us consider first the case of one square superconductor in a state fully penetrated by perpendicular magnetic flux. 
The left panel of Fig.~\ref{fig:Fig1} shows a magneto-optical image of the flux distribution in such a superconductor.
The most distinct feature here is the pair of dark diagonals across the sample.
Those lines, which indicate strong local magnetic shielding, are formed by abrupt in-plane turning of the induced critical current as the flow adapts to the shape of the sample.
The streamlines of this current flow are illustrated  in the right panel of the figure. 

\subsection{Two joined squares}

\begin{figure}[b]
  \centering
  \includegraphics[width=8.8cm]{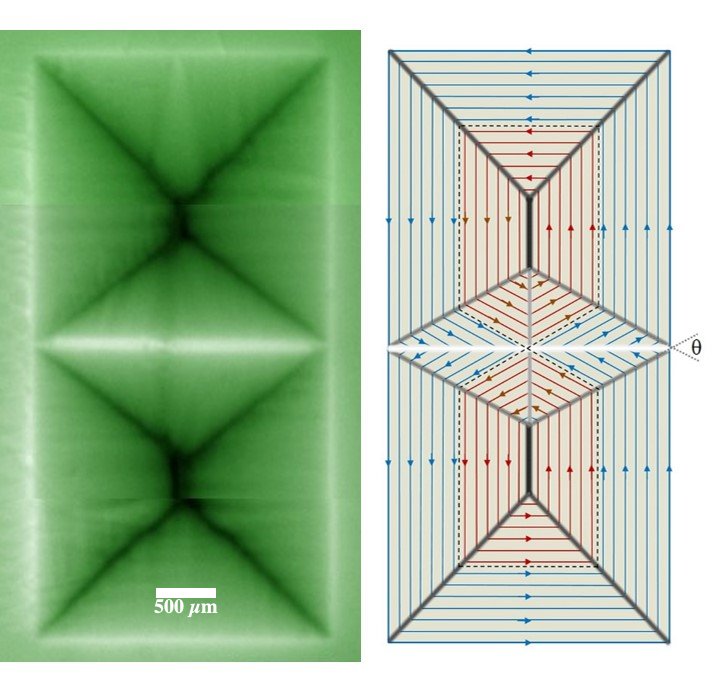} 
  \caption{
  Left: Magneto-optical image of the flux distribution in a superconducting film with a weak link dividing the rectangular sample into 2 squares. 
  Right: Streamlines  of induced flow of critical current drawn according to the Bean model for a sample of this geometry.
  Included as thick dark  lines are the corresponding d-lines, while the weak link appears bright.}
  \label{fig:Fig2}
\end{figure}

Next, let us consider two square superconductors joined to form a rectangle, where the joining interface has a reduced  critical current density, $J_i < J_c$. 
Shown in Fig.~\ref{fig:Fig2} (left) is a magneto-optical image of a superconductor with those characteristics.
The sample here is a Nb film shaped as a rectangle of aspect ratio 2:1, with a horizontal groove divides it into 2 squares connected by a weak link.
The image was recorded after cooling the sample in zero field, and then applying a perpendicular field of magnitude 150~Oe.

Again, the sample contour appears bright, but here an even brighter line is seen where the weak link is located. 
The full set of d-lines in this image can be understood from the current streamline pattern shown in the right panel of Fig.~\ref{fig:Fig2}.

The 4 dark d-lines extending from the sample corners, are similar to those seen in Fig.~\ref{fig:Fig1}. 
New in Fig.~\ref{fig:Fig2} is the set of d-lines forming a rhombus around the weak link. 
The weak link itself appears  bright since when the current flows across, all the  streamlines make a sharp clockwise turn, thus, adding a local magnetic  field in the same direction as the applied one.

When the current flows through the d-lines where the rhombus is, all the streamlines change direction in the counter-clockwise manner, thus creating local fields directed opposite to the applied one.
Hence, these d-lines appear dark in the images.
However, the turning of the current is here less than 90 degrees, and in MOI the d-lines forming the rhombus are therefore less dark than those extending from the corners.

Note that the figure also has a vertical and weakly dark d-line inside the rhombus. 
That d-line is created where the current rotates a relatively small angle in the counter-clockwise direction.
Finally, a pair of very distinct vertical d-lines are formed near the center of each square, at the boundary between current domains with $J_c$ flowing in opposite directions.

From the drawing in Fig.~\ref{fig:Fig2}~(right) it follows that there are two types of current streamlines, namely ($i$) those extending into both squares, and ($ii$) those localized within each square.
The boundary between the two types of flow is indicated by the dashed line.

A main reason to focus on d-lines is that from their highly visible geometrical configuration one can quantitatively measure the electric connectivity of joining interfaces~\cite{,Polyankii_magneto_1996,johansen_transparency_2019}.
In particular, the angle $\theta$ characterizing the rhomboid shape is related to the connectivity between 
the 2 squares by the simple equation,
\begin{equation}\label{eq:01}
J_i /J_c = \cos \theta.
\end{equation}
\noindent
A derivation of Eq.~\eqref{eq:01} is found in the Appendix.
Moreover, this relation applies also more generally, as shown below.

\subsection{Four joined squares}
\begin{figure}[t]
  \centering
  \includegraphics[width=5.8cm]{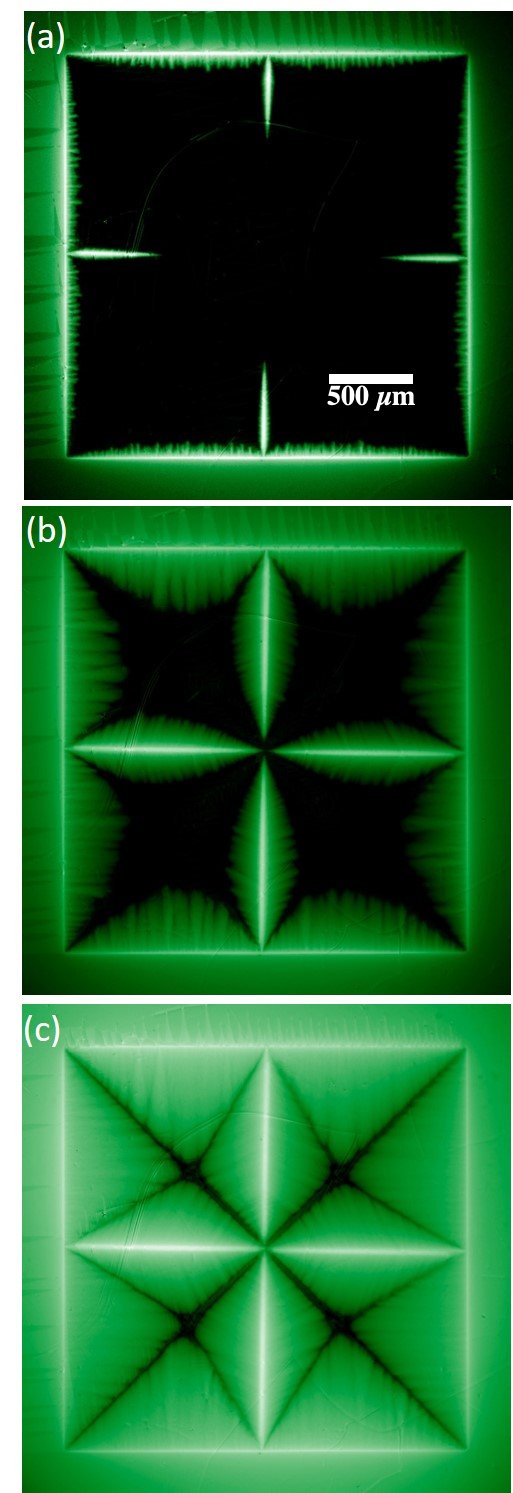} 
  \caption{Magneto-optical images of a Nb film consisting of 4 square superconductors connected by weak links (grooves in the film). 
  A perpendicular magnetic field was here increased from zero to 30~Oe (a), then to 80~Oe (b), and finally to 180~Oe (c). 
  The temperature was 7~K.}
\label{fig:Fig3}
\end{figure}

We now consider the case where 4 superconducting square plates are joined to form a larger square with 4 weak-link  connections.
The sample used for MOI investigation was prepared with a pair of perpendicular grooves to realize the 4 joining interfaces.
Presented in Fig.~\ref{fig:Fig3} are magneto-optical images of the flux penetration into this sample during gradual increase of a perpendicular magnetic field applied after initial zero-field cooling to 7~K.

The image in panel~(a) shows the flux density distribution when the applied field reached 30~Oe.
The flux pattern is here dominated by a needle-like penetration from each side of the square, exactly where the grooves are located.
The depth of the 1-D penetration is essentially the same from all the 4 sides, indicating that the grooves are weak links with essentially equal characteristics.

When the applied field is increased to 80~Oe, see panel~(b), the flux penetration into the weak links is nearly complete, while from the sample edges the penetration is relatively shallow.
Increasing further the field to 180 Oe, see panel~(c), the pattern of d-lines begins to stand out clearly in the whole sample.
Note that this d-line pattern is not a trivial extension of that in the case of 2 squares connected by a weak link. It is the result of a globally different shielding current flow pattern.

\begin{figure}[t]
  \centering
 \includegraphics[width=6.2cm]{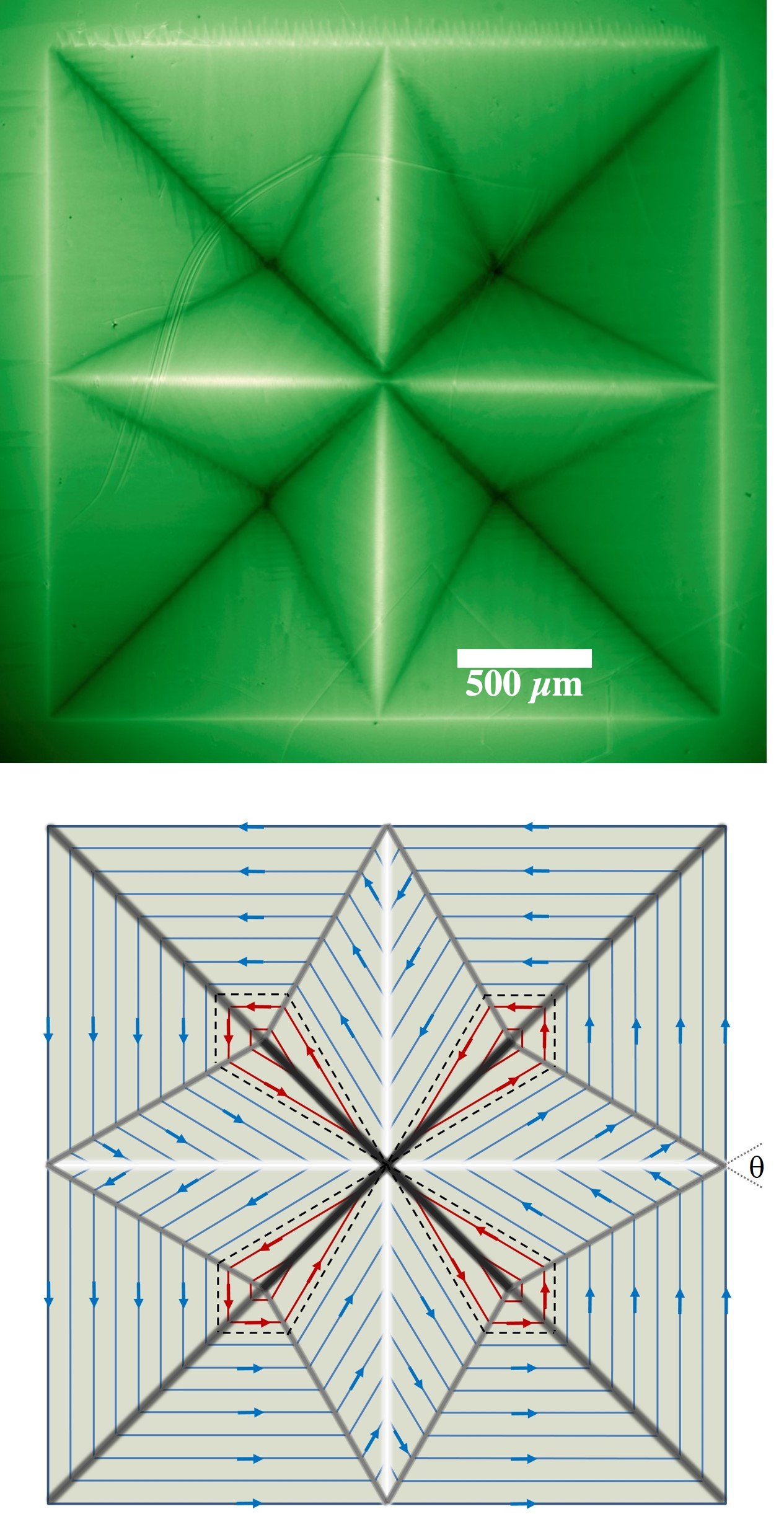}
  \caption{(Top) Magneto-optical image of the sample in an applied perpendicular field of 350~Oe, which creates full flux penetration in the sample. 
  (Bottom) Drawing of streamlines of the induced critical current flow, together with the corresponding set of d-lines.}
    \label{fig:4}
\end{figure}

To see the d-lines more fully developed, Fig.~\ref{fig:4}~(upper) shows the flux density distribution at 7~K after the applied perpendicular field was increased to 350~Oe.
Based on this set of d-lines one can draw the streamlines of the induced critical current.
The result is presented in the lower panel of Fig.~\ref{fig:4}. 
Included there are also the expected d-lines drawn in shades of gray.
Their darkness reflects the magnitude of reorientation in the current flow.
The correspondence between the d-lines in these 2 panels is striking.

\section{Discussion}
From the lower panel of Fig.~\ref{fig:4} one sees again two types of streamlines, namely, ($i$) in an outer region of the sample, a large number of streamlines pass through all 4 quadrants, while ($ii$) in 4 inner relatively small regions around the diagonals, the current flows in quadrilateral (kite-shaped) loops.
The boundary between the two types of current flow is marked in Fig.~\ref{fig:4} by the dashed lines.
A distinct feature of this flow pattern is, that from the midpoint of each edge a pair of d-lines form a V-shape quite similar to that seen in Fig.~\ref{fig:Fig2}.
From the detailed streamline pattern one finds that the angle, $\theta$,  between the d-lines is also here related to the weak link connectivity according to Eq.~\eqref{eq:01}, see the Appendix.

\begin{figure}[t]
  \centering
 \includegraphics[width=8.6cm]{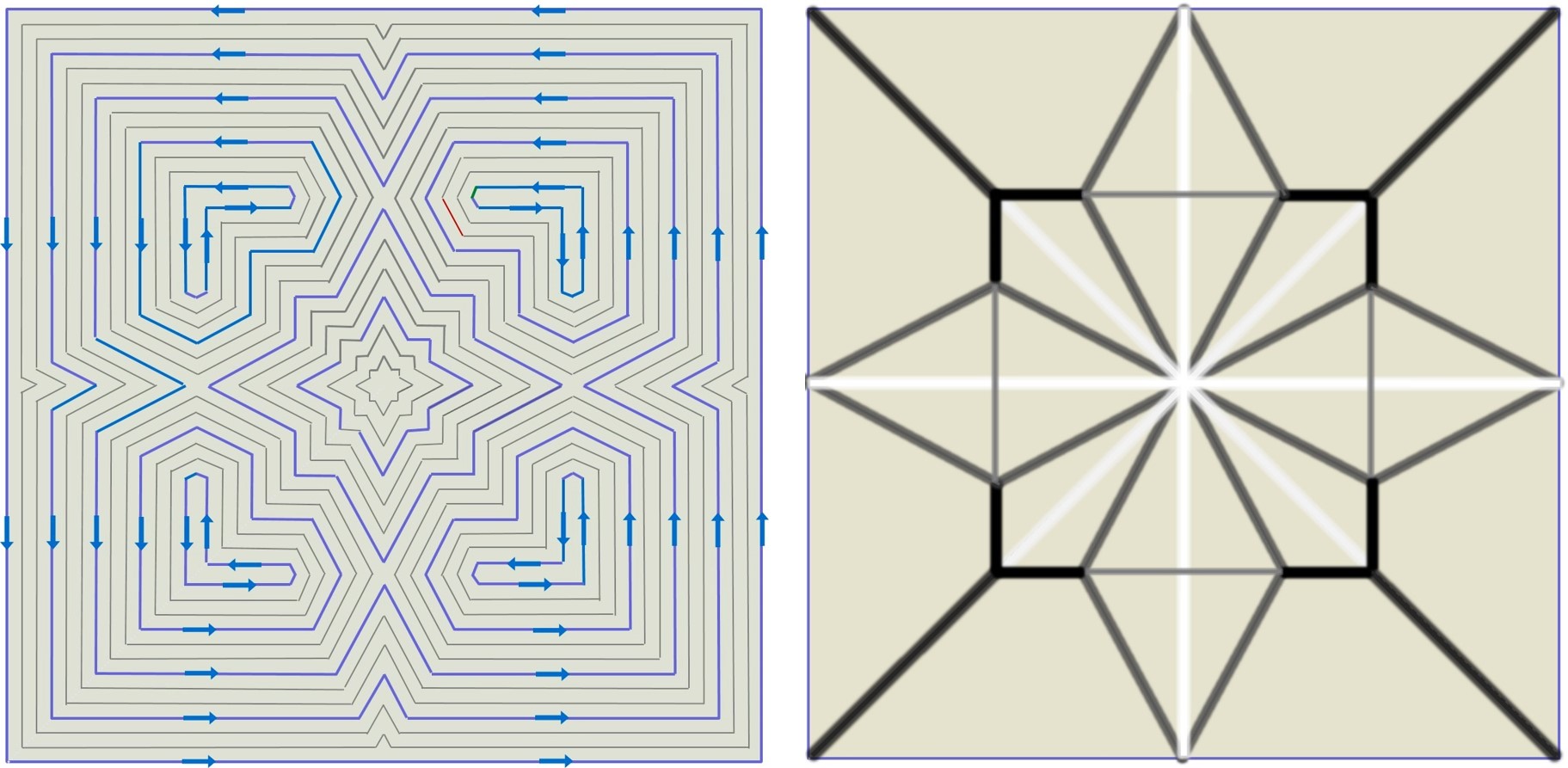}
  \caption{(Left) An a priori plausible current streamline pattern in a system of 4 jointed square superconducting plates connected by weak links, and (Right), the corresponding set of d-lines. Both turn out to be incorrect, see the main text.} 
\label{fig:5}
\end{figure}

The present observations show also that the flow pattern of the critical current in a system of 4 joined square plates is not a self-evident extension of the flow pattern in two joined plates shown  in Fig.~\ref{fig:Fig2}. 
In fact, without the d-lines being revealed one could anticipate other current flow patterns, e.g., the one shown in the left panel of Fig.~\ref{fig:5}.
That alternative flow pattern would generate the set of d-lines drawn in the right  panel, and is clearly incompatible with the present MOI observations. A similar observation was done by Schuster \textit {et al.} \cite{Schuster_flux_1996}, who concluded that the current pattern can be derived at convex corners by knowing the d-line formations.

To understand Nature's choice between the two candidate current flow patterns, energy considerations can be applied. For a superconducting film in a fully developed Bean critical state, the energy density of the current flow, being proportional to $J_c^2$ everywhere, will be the same irrespective of the flow pattern. 
Thus, the actual flow of  $J_c$ will be that having the  lowest energy associated with the magnetic field. 
That energy has a density proportional to $B^2$, suggesting to focus on the patterns of d-lines shown in Figs.~\ref{fig:4} and~\ref{fig:5}.

Comparing these sets of d-lines it is evident that the main differences are found along the sample diagonals.  
In Fig.~\ref{fig:5}~(right) a large part of the diagonals are bright, indicating  high flux density, whereas in Fig.~\ref{fig:4} the diagonals are dark  everywhere.
Hence, Nature's choice is the pattern with low flux density everywhere along the diagonals. 

In general, the current distribution is established as a result of dissipation processes. Therefore, the comparison of the energies corresponding to different distributions may be insufficient. That makes the stability of a current distribution much more involved, see, e.g., \cite{vestgarden_nucleation_2018} where the thermomagnetic instability is considered. Such an analysis falls beyond the scope of the present work

\section{Summary}
The present work has demonstrated an approach to reveal flow patterns of the critical current in homogeneous square superconducting plates connected by weak links.
The approach is based on observation of lines where the supercurrent abruptly changes the direction of flow when adapting to the configuration of weak links in the system.
It was shown using MOI that detection of flux density distributions, and in particular the distinct lines where the supercurrent abruptly changes direction of flow, allows to measure the electrical connectivity between 4 square plates forming a larger square.

\section*{acknowledgments}

The Nb films were grown in Laborat\'{o}rio de Conforma\c{c}\~{a}o Nanom\'{e}trica (IF-UFRGS), and the lithography was made in Laborat\'{o}rio de Microfabrica\c{c}\~{a}o (LNNano/CNPEM). 
The work was partially supported by the Sao Paulo Research Foundation (FAPESP) Grant No. 2017/24.786-4,
the Brazilian National Council for Scientific and Technological Development (CNPq), the Brazilian program Science without Borders, as well as the CAPES-SIU-2013/10046 project.

\section*{Derivation of Eq. (\ref{eq:01})}

\begin{figure}[b!]
  \centering
 \includegraphics[width=6.0cm]{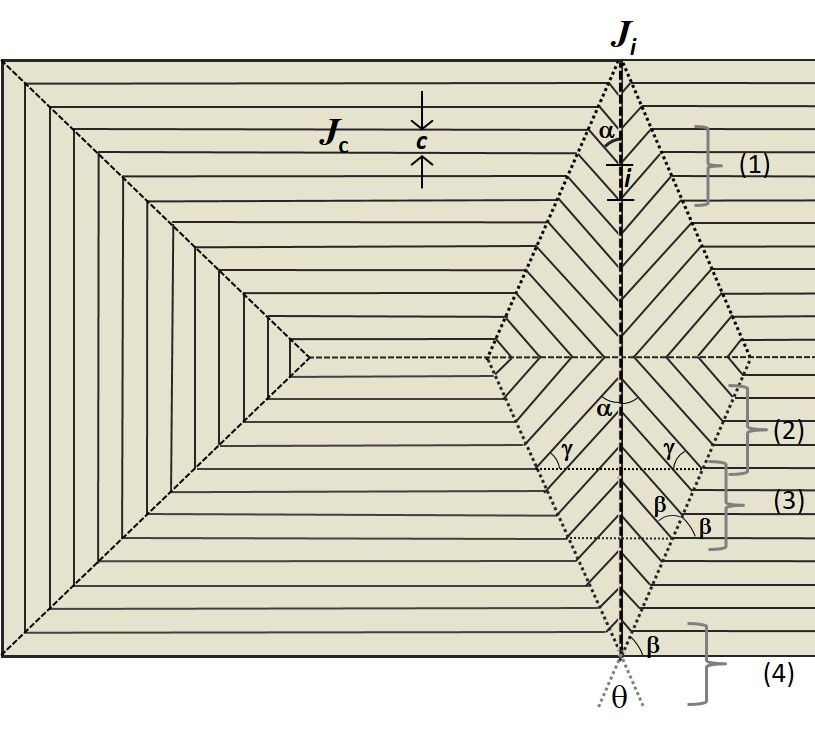}
  \caption{Streamlines of the critical current flow in a rectangular superconducting plate where a line with reduced critical  current $J_i < J_c$ divides the sample in 2 parts.} 
\label{fig:6}
\end{figure}

To see the validity of Eq.~\ref{eq:01} consider Fig.~\ref{fig:6}, which shows the streamline pattern of the current near a weak link extending across a rectangular superconductor~\cite{brandt_square_1995, schuster_flux_1997}.
Let the spacing of the  streamlines be constant, and denoted $c$ everywhere, except where the current crosses the weak link, where the spacing is $i > c$.

From the transparency definition \cite{johansen_transparency_2019} and the drawing in Fig. 6, it follows that the transparency ($\tau$) of the weak link in the region marked (1) is given by
\begin{equation}\label{eq:A1}
 \tau \equiv J_i/J_c = c/i = \sin \alpha.
\end{equation}
From the region marked (2) one sees that 
\begin{equation}\label{eq:A2}
\alpha + \gamma = 90^{\circ}, 
\end{equation}
and from region (3) that
\begin{equation}
\gamma + 2 \beta = 180^{\circ}.
\end{equation}
Finally, from region (4) one has
\begin{equation}
\theta + 2 \beta = 180^{\circ},
\end{equation}  
and thus, $\gamma = \theta$.
Then, Eq.~\eqref{eq:01} follows from Eqs.~\eqref{eq:A1} and~\eqref{eq:A2}.

\bibliography{4squareFIB-REF}

\begin{thebibliography}{15}%
\makeatletter
\providecommand \@ifxundefined [1]{%
 \@ifx{#1\undefined}
}%
\providecommand \@ifnum [1]{%
 \ifnum #1\expandafter \@firstoftwo
 \else \expandafter \@secondoftwo
 \fi
}%
\providecommand \@ifx [1]{%
 \ifx #1\expandafter \@firstoftwo
 \else \expandafter \@secondoftwo
 \fi
}%
\providecommand \natexlab [1]{#1}%
\providecommand \enquote  [1]{``#1''}%
\providecommand \bibnamefont  [1]{#1}%
\providecommand \bibfnamefont [1]{#1}%
\providecommand \citenamefont [1]{#1}%
\providecommand \href@noop [0]{\@secondoftwo}%
\providecommand \href [0]{\begingroup \@sanitize@url \@href}%
\providecommand \@href[1]{\@@startlink{#1}\@@href}%
\providecommand \@@href[1]{\endgroup#1\@@endlink}%
\providecommand \@sanitize@url [0]{\catcode `\\12\catcode `\$12\catcode
  `\&12\catcode `\#12\catcode `\^12\catcode `\_12\catcode `\%12\relax}%
\providecommand \@@startlink[1]{}%
\providecommand \@@endlink[0]{}%
\providecommand \url  [0]{\begingroup\@sanitize@url \@url }%
\providecommand \@url [1]{\endgroup\@href {#1}{\urlprefix }}%
\providecommand \urlprefix  [0]{URL }%
\providecommand \Eprint [0]{\href }%
\providecommand \doibase [0]{http://dx.doi.org/}%
\providecommand \selectlanguage [0]{\@gobble}%
\providecommand \bibinfo  [0]{\@secondoftwo}%
\providecommand \bibfield  [0]{\@secondoftwo}%
\providecommand \translation [1]{[#1]}%
\providecommand \BibitemOpen [0]{}%
\providecommand \bibitemStop [0]{}%
\providecommand \bibitemNoStop [0]{.\EOS\space}%
\providecommand \EOS [0]{\spacefactor3000\relax}%
\providecommand \BibitemShut  [1]{\csname bibitem#1\endcsname}%
\let\auto@bib@innerbib\@empty
%</preamble>
\bibitem [{\citenamefont {Durrell}\ \emph {et~al.}(2018)\citenamefont
  {Durrell}, \citenamefont {Ainslie}, \citenamefont {Zhou}, \citenamefont
  {Vanderbemden}, \citenamefont {Bradshaw}, \citenamefont {Speller},
  \citenamefont {Filipenko},\ and\ \citenamefont
  {Cardwell}}]{durrell_bulk_2018}%
  \BibitemOpen
  \bibfield  {author} {\bibinfo {author} {\bibfnamefont {J.~H.}\ \bibnamefont
  {Durrell}}, \bibinfo {author} {\bibfnamefont {M.~D.}\ \bibnamefont
  {Ainslie}}, \bibinfo {author} {\bibfnamefont {D.}~\bibnamefont {Zhou}},
  \bibinfo {author} {\bibfnamefont {P.}~\bibnamefont {Vanderbemden}}, \bibinfo
  {author} {\bibfnamefont {T.}~\bibnamefont {Bradshaw}}, \bibinfo {author}
  {\bibfnamefont {S.}~\bibnamefont {Speller}}, \bibinfo {author} {\bibfnamefont
  {M.}~\bibnamefont {Filipenko}}, \ and\ \bibinfo {author} {\bibfnamefont
  {D.~A.}\ \bibnamefont {Cardwell}},\ }\href {\doibase
  10.1088/1361-6668/aad7ce} {\bibfield  {journal} {\bibinfo  {journal}
  {Superconductor Science and Technology}\ }\textbf {\bibinfo {volume} {31}},\
  \bibinfo {pages} {103501} (\bibinfo {year} {2018})}\BibitemShut {NoStop}%
\bibitem [{\citenamefont
  {Johansen}(2000)}]{johansen_flux-pinning-induced_2000}%
  \BibitemOpen
  \bibfield  {author} {\bibinfo {author} {\bibfnamefont {T.~H.}\ \bibnamefont
  {Johansen}},\ }\href {\doibase 10.1088/0953-2048/13/10/201} {\bibfield
  {journal} {\bibinfo  {journal} {Superconductor Science and Technology}\
  }\textbf {\bibinfo {volume} {13}},\ \bibinfo {pages} {R121} (\bibinfo {year}
  {2000})}\BibitemShut {NoStop}%
\bibitem [{\citenamefont {Kordyuk}\ \emph {et~al.}(2001)\citenamefont
  {Kordyuk}, \citenamefont {Nemoshkalenko}, \citenamefont {Plyushchay},
  \citenamefont {Prikhna},\ and\ \citenamefont
  {Gawalek}}]{kordyuk_simple_2001}%
  \BibitemOpen
  \bibfield  {author} {\bibinfo {author} {\bibfnamefont {A.~A.}\ \bibnamefont
  {Kordyuk}}, \bibinfo {author} {\bibfnamefont {V.~V.}\ \bibnamefont
  {Nemoshkalenko}}, \bibinfo {author} {\bibfnamefont {A.~I.}\ \bibnamefont
  {Plyushchay}}, \bibinfo {author} {\bibfnamefont {T.~A.}\ \bibnamefont
  {Prikhna}}, \ and\ \bibinfo {author} {\bibfnamefont {W.}~\bibnamefont
  {Gawalek}},\ }\href {\doibase 10.1088/0953-2048/14/10/101} {\bibfield
  {journal} {\bibinfo  {journal} {Superconductor Science and Technology}\
  }\textbf {\bibinfo {volume} {14}},\ \bibinfo {pages} {L41} (\bibinfo {year}
  {2001})}\BibitemShut {NoStop}%
\bibitem [{\citenamefont {Bending}(1999)}]{bending_local_1999}%
  \BibitemOpen
  \bibfield  {author} {\bibinfo {author} {\bibfnamefont {S.~J.}\ \bibnamefont
  {Bending}},\ }\href {\doibase 10.1080/000187399243437} {\bibfield  {journal}
  {\bibinfo  {journal} {Advances in Physics}\ }\textbf {\bibinfo {volume}
  {48}},\ \bibinfo {pages} {449} (\bibinfo {year} {1999})}\BibitemShut
  {NoStop}%
\bibitem [{\citenamefont {Helseth}\ \emph {et~al.}(2002)\citenamefont
  {Helseth}, \citenamefont {Solovyev}, \citenamefont {Hansen}, \citenamefont
  {Il’yashenko}, \citenamefont {Baziljevich},\ and\ \citenamefont
  {Johansen}}]{helseth_faraday_2002}%
  \BibitemOpen
  \bibfield  {author} {\bibinfo {author} {\bibfnamefont {L.~E.}\ \bibnamefont
  {Helseth}}, \bibinfo {author} {\bibfnamefont {A.~G.}\ \bibnamefont
  {Solovyev}}, \bibinfo {author} {\bibfnamefont {R.~W.}\ \bibnamefont
  {Hansen}}, \bibinfo {author} {\bibfnamefont {E.~I.}\ \bibnamefont
  {Il’yashenko}}, \bibinfo {author} {\bibfnamefont {M.}~\bibnamefont
  {Baziljevich}}, \ and\ \bibinfo {author} {\bibfnamefont {T.~H.}\ \bibnamefont
  {Johansen}},\ }\href {\doibase 10.1103/PhysRevB.66.064405} {\bibfield
  {journal} {\bibinfo  {journal} {Physical Review B}\ }\textbf {\bibinfo
  {volume} {66}},\ \bibinfo {pages} {064405} (\bibinfo {year}
  {2002})}\BibitemShut {NoStop}%
\bibitem [{\citenamefont {Vlasko-Vlasov}\ \emph {et~al.}(1999)\citenamefont
  {Vlasko-Vlasov}, \citenamefont {Welp}, \citenamefont {Crabtree},\ and\
  \citenamefont {Mikheenko}}]{vlasko-vlasov_magneto-optical_1999}%
  \BibitemOpen
  \bibfield  {author} {\bibinfo {author} {\bibfnamefont {V.~K.}\ \bibnamefont
  {Vlasko-Vlasov}}, \bibinfo {author} {\bibfnamefont {U.}~\bibnamefont {Welp}},
  \bibinfo {author} {\bibfnamefont {G.~W.}\ \bibnamefont {Crabtree}}, \ and\
  \bibinfo {author} {\bibfnamefont {P.}~\bibnamefont {Mikheenko}},\ }\href@noop
  {} {\bibfield  {journal} {\bibinfo  {journal} {Physics and materials science
  of vortex states, flux pinning and dynamics}\ }\bibinfo {series} {{NATO}
  {ASI} series, {Series} {E}: {Applied} sciences},\ \textbf {\bibinfo {volume}
  {356}},\ \bibinfo {pages} {205} (\bibinfo {year} {1999})}\BibitemShut
  {NoStop}%
\bibitem [{\citenamefont {Altshuler}\ \emph {et~al.}(2004)\citenamefont
  {Altshuler}, \citenamefont {Johansen}, \citenamefont {Paltiel}, \citenamefont
  {Jin}, \citenamefont {Bassler}, \citenamefont {O~Ramos}, \citenamefont
  {Chen}, \citenamefont {Reiter}, \citenamefont {Zeldov},\ and\ \citenamefont
  {Chu}}]{altshuler_2004}%
  \BibitemOpen
  \bibfield  {author} {\bibinfo {author} {\bibfnamefont {E.}~\bibnamefont
  {Altshuler}}, \bibinfo {author} {\bibfnamefont {T.~H.}\ \bibnamefont
  {Johansen}}, \bibinfo {author} {\bibfnamefont {Y.}~\bibnamefont {Paltiel}},
  \bibinfo {author} {\bibfnamefont {P.}~\bibnamefont {Jin}}, \bibinfo {author}
  {\bibfnamefont {K.}~\bibnamefont {Bassler}}, \bibinfo {author} {\bibfnamefont
  {O.}~\bibnamefont {O~Ramos}}, \bibinfo {author} {\bibfnamefont
  {Q.}~\bibnamefont {Chen}}, \bibinfo {author} {\bibfnamefont {G.}~\bibnamefont
  {Reiter}}, \bibinfo {author} {\bibfnamefont {E.}~\bibnamefont {Zeldov}}, \
  and\ \bibinfo {author} {\bibfnamefont {C.}~\bibnamefont {Chu}},\ }\href
  {\doibase 10.1088/0953-2048/13/10/201} {\bibfield  {journal} {\bibinfo
  {journal} {Physical Review B}\ }\textbf {\bibinfo {volume} {70}},\ \bibinfo
  {pages} {140505} (\bibinfo {year} {2004})}\BibitemShut {NoStop}%
\bibitem [{\citenamefont {Yurchenko}\ \emph {et~al.}(2009)\citenamefont
  {Yurchenko}, \citenamefont {Johansen},\ and\ \citenamefont
  {Galperin}}]{yurchenko_dendritic_2009}%
  \BibitemOpen
  \bibfield  {author} {\bibinfo {author} {\bibfnamefont {V.~V.}\ \bibnamefont
  {Yurchenko}}, \bibinfo {author} {\bibfnamefont {T.~H.}\ \bibnamefont
  {Johansen}}, \ and\ \bibinfo {author} {\bibfnamefont {Y.~M.}\ \bibnamefont
  {Galperin}},\ }\href {\doibase 10.1063/1.3224713} {\bibfield  {journal}
  {\bibinfo  {journal} {Low Temperature Physics}\ }\textbf {\bibinfo {volume}
  {35}},\ \bibinfo {pages} {619} (\bibinfo {year} {2009})}\BibitemShut
  {NoStop}%
\bibitem [{\citenamefont {Colauto}\ \emph {et~al.}(2008)\citenamefont
  {Colauto}, \citenamefont {Patino}, \citenamefont {Blamire},\ and\
  \citenamefont {Ortiz}}]{colauto_boundaries_2008}%
  \BibitemOpen
  \bibfield  {author} {\bibinfo {author} {\bibfnamefont {F.}~\bibnamefont
  {Colauto}}, \bibinfo {author} {\bibfnamefont {E.~J.}\ \bibnamefont {Patino}},
  \bibinfo {author} {\bibfnamefont {M.~G.}\ \bibnamefont {Blamire}}, \ and\
  \bibinfo {author} {\bibfnamefont {W.~A.}\ \bibnamefont {Ortiz}},\ }\href
  {\doibase 10.1088/0953-2048/21/4/045018} {\bibfield  {journal} {\bibinfo
  {journal} {Superconductor Science \& Technology}\ }\textbf {\bibinfo {volume}
  {21}},\ \bibinfo {pages} {045018} (\bibinfo {year} {2008})}\BibitemShut
  {NoStop}%
\bibitem [{\citenamefont {Polyanskii}\ \emph {et~al.}(1996)\citenamefont
  {Polyanskii}, \citenamefont {Gurevich}, \citenamefont {Pashitski},
  \citenamefont {Heinig}, \citenamefont {Redwing}, \citenamefont {Nordman},\
  and\ \citenamefont {Larbalestier}}]{Polyankii_magneto_1996}%
  \BibitemOpen
  \bibfield  {author} {\bibinfo {author} {\bibfnamefont {A.~A.}\ \bibnamefont
  {Polyanskii}}, \bibinfo {author} {\bibfnamefont {A.}~\bibnamefont
  {Gurevich}}, \bibinfo {author} {\bibfnamefont {A.~E.}\ \bibnamefont
  {Pashitski}}, \bibinfo {author} {\bibfnamefont {N.~F.}\ \bibnamefont
  {Heinig}}, \bibinfo {author} {\bibfnamefont {R.~D.}\ \bibnamefont {Redwing}},
  \bibinfo {author} {\bibfnamefont {J.~E.}\ \bibnamefont {Nordman}}, \ and\
  \bibinfo {author} {\bibfnamefont {D.~C.}\ \bibnamefont {Larbalestier}},\
  }\href {\doibase 10.1103/PhysRevB.53.8687} {\bibfield  {journal} {\bibinfo
  {journal} {Phys. Rev. B}\ }\textbf {\bibinfo {volume} {53}},\ \bibinfo
  {pages} {8687} (\bibinfo {year} {1996})}\BibitemShut {NoStop}%
\bibitem [{\citenamefont {Johansen}\ \emph {et~al.}(2019)\citenamefont
  {Johansen}, \citenamefont {Colauto}, \citenamefont {de~Andrade},
  \citenamefont {Oliveira},\ and\ \citenamefont
  {Ortiz}}]{johansen_transparency_2019}%
  \BibitemOpen
  \bibfield  {author} {\bibinfo {author} {\bibfnamefont {T.~H.}\ \bibnamefont
  {Johansen}}, \bibinfo {author} {\bibfnamefont {F.}~\bibnamefont {Colauto}},
  \bibinfo {author} {\bibfnamefont {A.~M.~H.}\ \bibnamefont {de~Andrade}},
  \bibinfo {author} {\bibfnamefont {A.~A.~M.}\ \bibnamefont {Oliveira}}, \ and\
  \bibinfo {author} {\bibfnamefont {W.~A.}\ \bibnamefont {Ortiz}},\ }\href
  {\doibase 10.1109/TASC.2019.2899209} {\bibfield  {journal} {\bibinfo
  {journal} {IEEE Transactions on Applied Superconductivity}\ }\textbf
  {\bibinfo {volume} {29}},\ \bibinfo {pages} {8002304} (\bibinfo {year}
  {2019})}\BibitemShut {NoStop}%
\bibitem [{\citenamefont {Schuster}\ \emph {et~al.}(1996)\citenamefont
  {Schuster}, \citenamefont {Kuhn},\ and\ \citenamefont
  {Brandt}}]{Schuster_flux_1996}%
  \BibitemOpen
  \bibfield  {author} {\bibinfo {author} {\bibfnamefont {T.}~\bibnamefont
  {Schuster}}, \bibinfo {author} {\bibfnamefont {H.}~\bibnamefont {Kuhn}}, \
  and\ \bibinfo {author} {\bibfnamefont {E.~H.}\ \bibnamefont {Brandt}},\
  }\href {\doibase 10.1103/PhysRevB.54.3514} {\bibfield  {journal} {\bibinfo
  {journal} {Phys. Rev. B}\ }\textbf {\bibinfo {volume} {54}},\ \bibinfo
  {pages} {3514} (\bibinfo {year} {1996})}\BibitemShut {NoStop}%
\bibitem [{\citenamefont {Vestgarden}\ \emph {et~al.}(2018)\citenamefont
  {Vestgarden}, \citenamefont {Johansen},\ and\ \citenamefont
  {Galperin}}]{vestgarden_nucleation_2018}%
  \BibitemOpen
  \bibfield  {author} {\bibinfo {author} {\bibfnamefont {J.~I.}\ \bibnamefont
  {Vestgarden}}, \bibinfo {author} {\bibfnamefont {T.~H.}\ \bibnamefont
  {Johansen}}, \ and\ \bibinfo {author} {\bibfnamefont {Y.~M.}\ \bibnamefont
  {Galperin}},\ }\href {\doibase 10.1063/1.5037549} {\bibfield  {journal}
  {\bibinfo  {journal} {Low Temperature Physics}\ }\textbf {\bibinfo {volume}
  {44}},\ \bibinfo {pages} {460} (\bibinfo {year} {2018})}\BibitemShut
  {NoStop}%
\bibitem [{\citenamefont {Brandt}(1995)}]{brandt_square_1995}%
  \BibitemOpen
  \bibfield  {author} {\bibinfo {author} {\bibfnamefont {E.~H.}\ \bibnamefont
  {Brandt}},\ }\href {\doibase 10.1103/PhysRevLett.74.3025} {\bibfield
  {journal} {\bibinfo  {journal} {Physical Review Letters}\ }\textbf {\bibinfo
  {volume} {74}},\ \bibinfo {pages} {3025} (\bibinfo {year}
  {1995})}\BibitemShut {NoStop}%
\bibitem [{\citenamefont {Schuster}\ \emph {et~al.}(1997)\citenamefont
  {Schuster}, \citenamefont {Kuhn}, \citenamefont {Brandt},\ and\ \citenamefont
  {Klaumünzer}}]{schuster_flux_1997}%
  \BibitemOpen
  \bibfield  {author} {\bibinfo {author} {\bibfnamefont {T.}~\bibnamefont
  {Schuster}}, \bibinfo {author} {\bibfnamefont {H.}~\bibnamefont {Kuhn}},
  \bibinfo {author} {\bibfnamefont {E.~H.}\ \bibnamefont {Brandt}}, \ and\
  \bibinfo {author} {\bibfnamefont {S.}~\bibnamefont {Klaumünzer}},\ }\href
  {\doibase 10.1103/PhysRevB.56.3413} {\bibfield  {journal} {\bibinfo
  {journal} {Physical Review B}\ }\textbf {\bibinfo {volume} {56}},\ \bibinfo
  {pages} {3413} (\bibinfo {year} {1997})}\BibitemShut {NoStop}%
\end{thebibliography}%
\end{document}